\documentclass{jdmdh}
\usepackage{array}
\usepackage{pgfplots}
\pdfoutput=1

\title{Node similarity as a basic principle behind\\ connectivity in complex networks}
\author[1,2]{Matthias Scholz}
\affil[1]{Ernst-Moritz-Arndt-University, Greifswald, Germany} 
\affil[2]{Present address: University of Trento, Italy} 

\corrauthor{Matthias Scholz}{matthias.scholz@network-science.org}

\begin{document}

\maketitle

\abstract{How are people linked in a highly connected society? Since in many networks a
power-law (scale-free) node-degree distribution can be observed, power-law might be seen
as a universal characteristics of networks. But this study of communication in the Flickr
social online network reveals that power-law node-degree distributions are restricted to
only sparsely connected networks. More densely connected networks, by contrast, show an
increasing divergence from power-law. This work shows that this observation is consistent
with the classic idea from social sciences that similarity is the driving factor behind
communication in social networks. The strong relation between communication strength and
node similarity could be confirmed by analyzing the Flickr network. It also is shown that
node similarity as a network formation model can reproduce the characteristics of different
network densities and hence can be used as a model for describing the topological transition
from weakly to strongly connected societies.
}


\section{Introduction}

\strut
\vspace{-4ex}

\begin{figure}[!tpb]
  \centering
  \centerline{\includegraphics[width=0.51\linewidth]{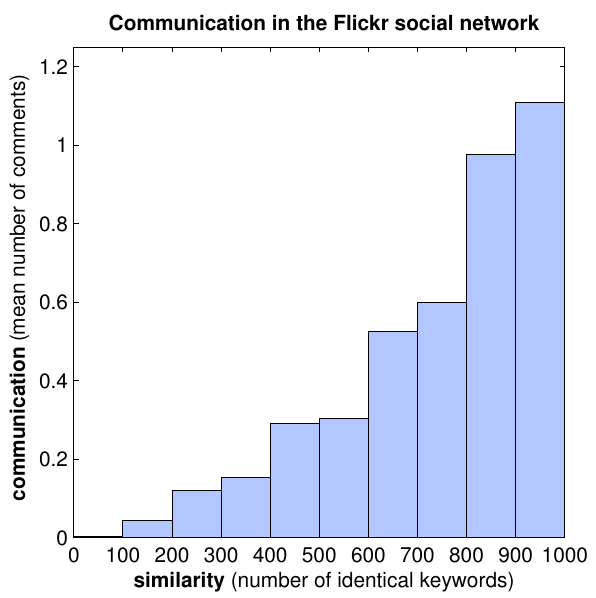}}
   \caption[abc]{
     Similarity and communication in the 
     Flickr{\scriptsize\textsuperscript{\textregistered}} 
     social network.
     For each pair of a randomly chosen set of 10,000 
     Flickr{\scriptsize\textsuperscript{\textregistered}} users, 
     the number of identically used keywords (tags) is set into 
     relation with the pair-wise communication strengths. 
     The histogram shows the mean communication strength of all pairs 
     within intervals of 100 keywords. 
     This clearly confirms that a higher number of identical 
     keywords is strongly related to higher communication.\\
     }\label{fig:tag_similarity}
\end{figure}

In an increasingly interconnected world, it must be of huge interest to understand the topology of a highly connected society;
important, for example, for predicting the spread of epidemic diseases~\citep{PasVes01}.
A basic measure to describe network topologies is the distribution of the number of links per network-node. 
Many real networks show a node-degree distribution that approximately follows a 
power-law --- a right-skewed heavy-tailed distribution 
also known as scale-free distribution~\citep{BarAlb99}.
But other real networks show a truncated power-law or even an exponentially shaped node-degree distribution \citep{AmaScaBar00,New01,JeoMasBar01}. 
\\ 
To investigate network topologies it is essential to understand the basic principles behind connectivity or, more precisely, the process of network formation.
Most of the current models are basically focused on reproducing a power-law (scale-free) network topology.
The most popular model is a network growth model based 
on the idea of preferential attachment: new nodes prefer to link to existing highly connected nodes \citep{Pri76,BarAlb99}.
But a high node-degree may rather be the result than the cause of connectivity as shown by other models of network formation,
including the node copying model \citep{KleKumRag99} and the fitness model \citep{CalCapLos02,SerCalBut04}.
Even though most models reproduce quite well a power-law distribution, they do not explain the frequently observed divergences from power-law.
\pagebreak[1]\\
Social sciences have a long history in explaining social communication and interaction 
and a huge amount of literature from this field 
suggests that similarity (homophily) is the major factor for connectivity in social networks as, for example, reviewed by 
%McPherson et al.~\citep{PheSmiCoo01}.
\citet{PheSmiCoo01}.
People tend to associate with those sharing similar interests, tastes,  beliefs, social backgrounds, and also similar popularity. This is often expressed by the adage `Birds of a feather flock together'.
\\
Recent analysis of mobile phone data further confirms that communication is strongly related to geographic distance. 
There is a higher chance of people calling each other if they live closer to each other (similar location). 
The total amount of communication between two cities depends on their distance and population size, 
which can be well described by a gravitation model \citep{LamBloKer08,KriCalRat09}.
\\
In biology, interactions between proteins or other molecules require an exact fit or complementarity of their complex surfaces
which have to be treated synonymously with similarity in the context of connectivity.
\\
For communication and interaction, space and time are often the dominant factors. `To be in the right place at the right time' works often as the basic principle for getting connected, but beside fitting in space and time additional properties are important: for instance similar surfaces of molecules, or similar interests of people. In mobile phone networks,  it can be shown that other factors besides geographic distance influence communication, e.g., 
language \citep{LamBloKer08}. Such additional factors become even more important in virtual communities in which geographic distance does not matter and written communication does not require the presence of the networked partner at the same time. 
\\
In information networks, location and time are also not the dominant factors. 
In general, articles are linked because of similar topics, 
scientific citations have a strong relevance to the author's work, and websites are mostly linked to websites 
of similar content \citep{FlaLawGil02}.
\pagebreak[2]\\
Online social networks are an ideal source for investigating complex networks because of the often huge number of users, their link and communication profiles, and the availability of additional metadata such as tags (keywords).
Several recent studies confirm the impact of similarity on links
in social online networks by analyzing tag (keyword) metadata between users 
\citep{MarNaaBoy06,SchBarCat10,AieBarCat10}.
But most studies focus on an unweighted contact (declared friends) network structure.
By contrast, this study analyzes communication strength between users. 
This provides us a more precise description of user interactions in terms of
weighted links or contact intensities useful for analyzing the transition from
sparsely connected to densely connected networks.
In the first step, by analyzing the Flickr{\scriptsize\textsuperscript{\textregistered}} social online network, this study shows that communication strength is directly related to tag  (keyword) similarity. In the second step, the Flickr{\scriptsize\textsuperscript{\textregistered}} network is used to analyze different network densities. It turns out that more densely connected networks show an increasing divergence from the power-law distribution. This characteristic can be reproduced by a network formation model based on similarity, as shown by the Euclidean distance model proposed in this work.

\section{Similarity in the Flickr network}
Flickr{\scriptsize\textsuperscript{\textregistered}} is
an online photo sharing community.
Here, we analyze how users interact and communicate by commenting on photos of other users.
Data were collected in 2009 by using the application programming interface (API) to the 
Flickr{\scriptsize\textsuperscript{\textregistered}} database at 
\linebreak[4] \href{https://www.flickr.com/services/api/}{\texttt{https://www.flickr.com/services/api/}}\,.
\\
The number of comments of one user A to another user B is used to define the strength of communication,
and hence gives the weight of the link between A and B. In this study, similarity between two users is based on keywords (tags) people use
to describe their photos. The similarity is defined as the number of identical tags of two users: the size of the
intersection of the tag sets of user A and B from all of their photos.
People who use the same keywords are supposed to have similar photographic interests which, in turn, may lead to communication.
\\
Setting the number of identical keywords into relation with the number of comments between two individuals, as shown in
Figure~\ref{fig:tag_similarity}, reveals a clear dependency between similarity and communication strength.
The intensity of communication between two individuals is strongly related to the number of identically used keywords,
thereby confirming empirically that communication strength depends on similar interest of individuals.

\begin{figure}[!tpb]
   \centerline{\includegraphics[width=1.0\linewidth]{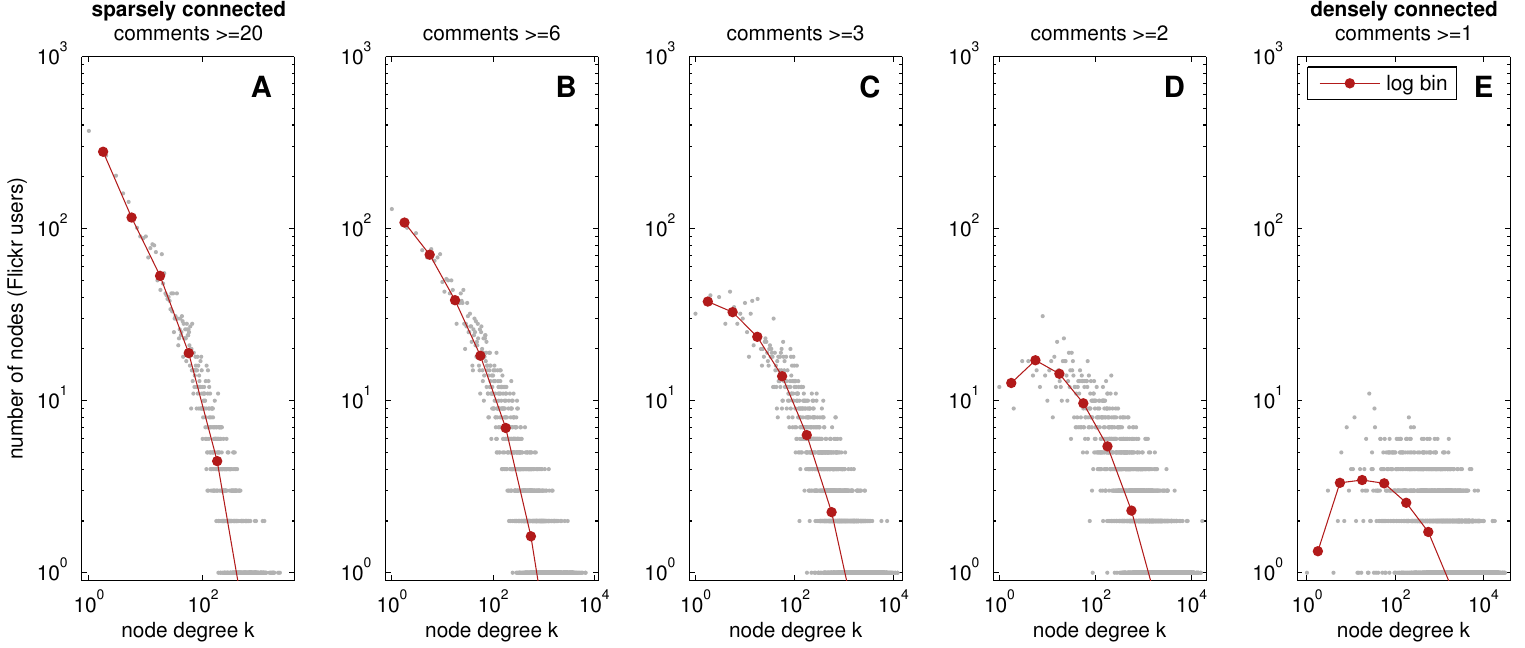}}
   \caption[abc]{
     Node-degree distributions of the  
     Flickr{\scriptsize\textsuperscript{\textregistered}} online network.
     Distributions of a densely connected Flickr-group are
     plotted for different connectivity levels on log-log scales.
     Logarithmic binning is used to show noise-reduced distributions (red line).
     Two individuals are defined to be linked when the number of their
     comments exceed a certain threshold. 
     Different thresholds lead to networks that differ in 
     their overall connectivity level.
     (\textsf{\textbf{A}}) Counting only strong links with more 
     than 20 comments leads to a sparsely connected network showing the 
     typical scale-free power-law distribution. 
     (\textsf{\textbf{B}}, \textsf{\textbf{C}}, and \textsf{\textbf{D}}) 
     Moderate thresholds lead to the often observed saturation effects 
     in lower node-degrees: 
     the number of nodes of low degree is smaller than expected 
     for a scale-free power-law topology.
     (\textsf{\textbf{E}}) A densely connected network (counting also 
     very weak links of only one comment) does not follow anymore
     a power-law.\\
     }\label{fig:flickr_node_degree_distribution_comments}
 \end{figure}

\begin{figure}[!tpb]
  \centerline{\includegraphics[width=1.0\linewidth]{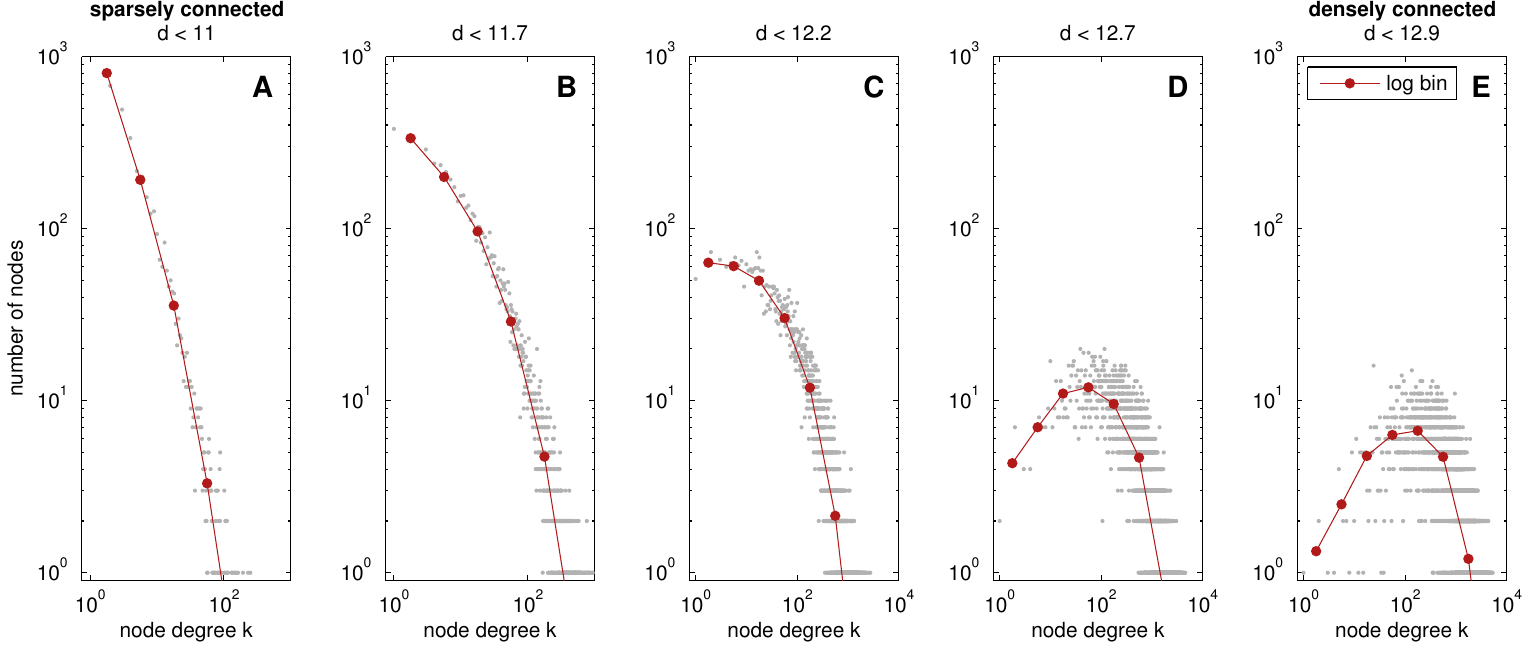}}
   \caption[abc]{
     The similarity model. 
     Based on a random data set, two nodes are defined as connected (similar)
     when their Euclidean distance $d$ is below a certain threshold.
     The distributions, plotted on log-log scales, depend on network density.
     (\textsf{\textbf{A}}) With a strong threshold only very similar 
     nodes are connected. 
     This represents a sparsely connected network showing the 
     typical scale-free power-law like distribution.
     \mbox{(\textsf{\textbf{B-E}})} In increasingly connected networks as 
     given by weaker thresholds, the number 
     of nodes having a small degree decreases.
     A similarity model is able to reproduce the diversity of 
     distributions, from sparsely to densely connected networks, 
     as found in the real 
     network (Figure~\ref{fig:flickr_node_degree_distribution_comments}).\\
     }\label{fig:paper_euclid_loglog_plot}
 \end{figure}

\section{From sparse to dense networks}
In order to investigate how node-degree distributions depend on network density, 
the difference between sparsely and densely connected topologies is analyzed. 
Since most networks are rather sparsely connected, 
including the Flickr{\scriptsize\textsuperscript{\textregistered}} network as a whole,
a more densely connected subset of Flickr{\scriptsize\textsuperscript{\textregistered}} is exemplarily
chosen: the Flickr-group `Light Painters Society' (id:1066685@N25) having 6,036 members (nodes). 
By using different thresholds for the number of comments to be accepted as a link, the degree of overall connectivity can be varied from sparsely to densely 
connected networks.
\\
Figure~\ref{fig:flickr_node_degree_distribution_comments} shows the in-degree distribution counting
only strong links (more than or equal to 20~comments), medium-weighted links (more than or equal to 2,~3, or 6~comments), 
and all, including very weak links (at least one comment).
It reveals that only a sparsely connected network shows the typical scale-free power-law like distribution.
Densely connected networks, by contrast, show a distribution which is very distinct from power-law.

\section*{The node similarity model}
The observed characteristics of real networks can be reproduced by a simple similarity model based on Euclidean distance in pure random data.
This is demonstrated by artificially generating a network from a 100\,x\,8000 normally distributed random data matrix~{$\mathcal X$}, according to $m=100$ properties and $N=8000$ network nodes. 
Two nodes $x_i$ and $x_j$ are defined as connected if their Euclidean 
distance $d = \parallel{x_i}-{x_j}\parallel$
is below a certain threshold.
Increasing this threshold means changing the network density from sparsely to densely connected.
As shown in Figure~\ref{fig:paper_euclid_loglog_plot}, a similarity model generates the same shapes of node-degree distributions as observed in the real network (Figure~\ref{fig:flickr_node_degree_distribution_comments}).
\\
A MATLAB\textsuperscript{\textregistered} implementation of 
the similarity model is available at:\\
\href{http://www.network-science.org/similaritymodel.html}
{\texttt{http://www.network-science.org/similarity\-model.html}}\,.

\begin{figure*}[!b]
   \centerline{\includegraphics[width=1.0\linewidth]{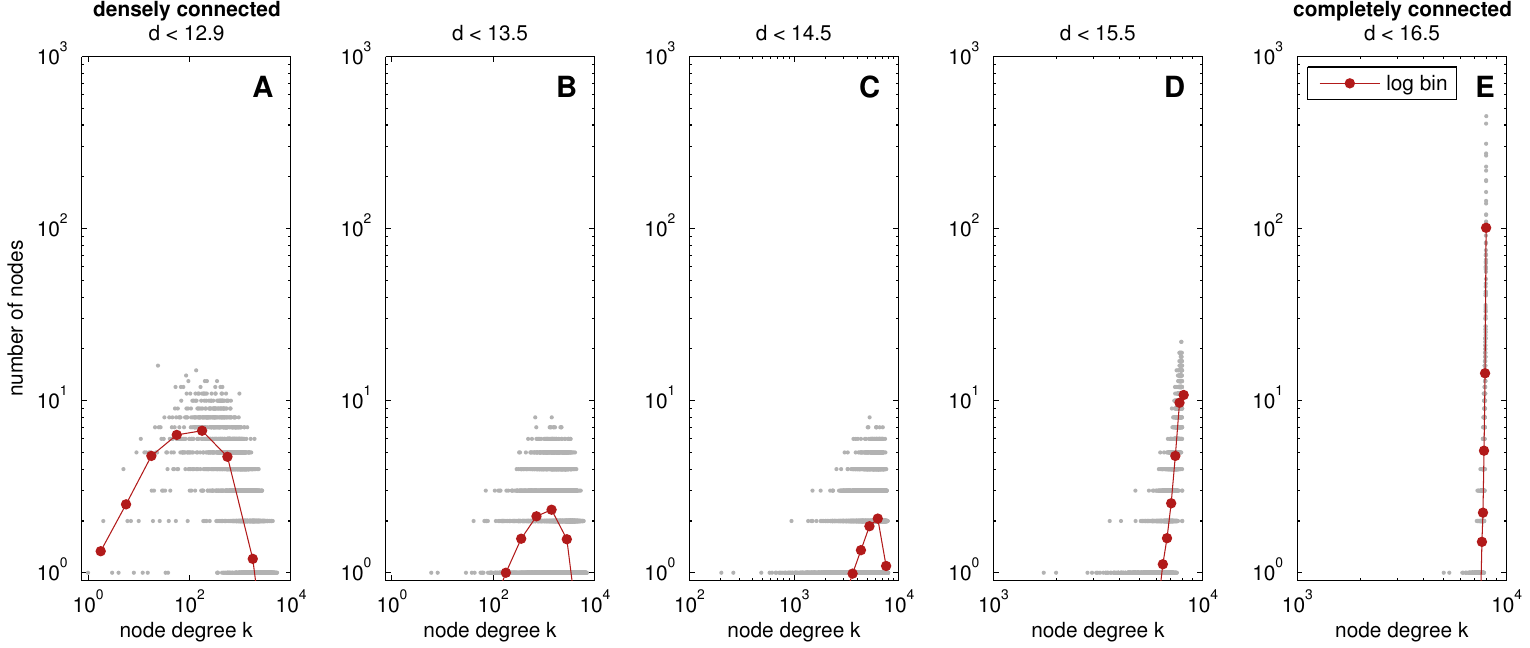}}
   \caption[abc]{
     Topological changes from densely to almost completely connected networks. 
     Plotted are the node-degree distributions of increasingly connected networks 
     generated by the similarity model (\textsf{\textbf{A-E}}).
     In almost completely connected networks (\textsf{\textbf{D}} 
     and \textsf{\textbf{E}}) 
     the node-degree distribution appears as an inverse power-law: 
     most nodes have a high degree whereas only few nodes have a low degree.   
     }\label{fig:paper_euclid_loglog_plot_complete}
 \end{figure*}

\section*{Benefits of a similarity based model}
Beside the relation shown between similarity and connectivity strength,
there are a number of other points that show that a similarity model is an appropriate and natural way to describe real complex networks:\\
a) Most network formation models are developed to reproduce only power-law distributions such as in Figure~\ref{fig:flickr_node_degree_distribution_comments}A. Thus, they cannot explain node-degree distributions distinct from a power-law as in Figure~\ref{fig:flickr_node_degree_distribution_comments}\,C to E. A similarity model, by contrast, covers naturally the full observed diversity from power-law to non-power-law distributions.\\
b) A similarity model does not depend on dynamics in network size such as an increase or decrease in the total number of networks nodes. It therefore works within situations of network growth as well as shrinkage or even for pure reorganization of links in a network of constant size. Since also power-law like distributions can be reproduced by the similarity model, the observed power-law characteristics of real networks are not necessarily a result of network growth.\\ 
c) Because of the usually undirected property of similarity, it is a natural model for undirected networks in which connections are induced from both sides as in social networks. But similarity works also in a directed manner when additional factors such as time in a growth model (e.g., citation network) enforce directed relations.\\
d) Similarity does not require global knowledge such as node-degree about all network nodes. Similarity refers only to the local environment of people in real physical as well as in virtual communication worlds. People who live in the same place, engage in similar activities, or members of online communities meet each other and connect according to their similar behaviors and interests. A global knowledge about all people is not necessary.\\ 
e) A similarity model explains the topological transition from sparsely to densely or even completely connected networks which a pure power-law model does not.
Completely connected networks in which each node is connected to each other do not follow a power-law distribution, instead, all $N$ nodes have the same maximum degree, $k=N-1$. Thus, with increasing connectivity there must be a transition from the power-law topologies (Fig.~\ref{fig:paper_euclid_loglog_plot}A) of sparsely connected networks 
to the peaked distributions (Fig.~\ref{fig:paper_euclid_loglog_plot_complete}E) of completely connected networks. 
A similarity model can describe such transition from sparsely to densely connected networks as shown in Figure~\ref{fig:paper_euclid_loglog_plot} and, in addition, to completely connected networks (Fig.~\ref{fig:paper_euclid_loglog_plot_complete}). 
For almost completely connected networks the similarity model predicts a left-skewed distribution inverse to the power-law in which most nodes have a high degree and only a few nodes have a low degree.

\section*{Conclusion}
This work demonstrates that the frequently observed scale-free power-law distribution can be well reproduced by a model which is purely based on the idea of node similarity. 
Since similarity is independent of dynamics in network size such as growth or shrinkage, 
the observed power-law of real networks is not necessarily caused by the growth of networks.
In addition, a similarity model shows that the frequently observed distributions distinct from power-law are a characteristics of more densely connected networks. 
This means that the differences we can observe in node-degree distributions of real networks 
are mainly given by their overall link density: whereas the typical sparsely connected networks show power-law distributions, densely connected networks show non-power-law distributions.
This can be further extended to almost completely connected networks as can be found in a family or a small village in which everyone knows everyone else. While in sparsely connected power-law networks most nodes have a low number of links and only a few are highly linked, almost completely connected networks show the opposite: most nodes have a high or even maximum degree and only a few nodes have lower degrees. These less connected nodes may represent outsiders in an almost completely connected clique. 
Since a similarity model explains the entire topological transition from sparsely to densely connected networks it is able to explain the transition from lowly connected to highly connected societies.

\section*{Acknowledgments}
Funding was provided by the German Ministry for Science and Research (BMBF) within the program Entrepreneurial Regions: Competence Centers, under code 03\,ZIK\,011. I~would like to thank the Department of Mathematics and Informatics of Ernst-Moritz-Arndt-University for providing computational facilities and Martin J.~Fraunholz for valuable discussions and critically reading the manuscript.

\pagebreak[4]
\bibliographystyle{plainnat}
% \bibliography{scholz}

\begin{thebibliography}{16}
\providecommand{\natexlab}[1]{#1}
\providecommand{\url}[1]{\texttt{#1}}
\expandafter\ifx\csname urlstyle\endcsname\relax
  \providecommand{\doi}[1]{doi: #1}\else
  \providecommand{\doi}{doi: \begingroup \urlstyle{rm}\Url}\fi

\bibitem[Aiello et~al.(2010)Aiello, Barrat, Cattuto, Ruffo, and
  Schifanella]{AieBarCat10}
L.~M. Aiello, A.~Barrat, C.~Cattuto, G.~Ruffo, and R.~Schifanella.
\newblock Link creation and profile alignment in the {aNobii} social network.
\newblock In \emph{SocialCom/PASSAT}, pages 249--256, 2010.
\newblock \doi{10.1109/SocialCom.2010.42}.

\bibitem[Amaral et~al.(2000)Amaral, Scala, Barth\'el\'emy, and
  Stanley]{AmaScaBar00}
L.~A.~N. Amaral, A.~Scala, M.~Barth\'el\'emy, and H.~E. Stanley.
\newblock Classes of small-world networks.
\newblock \emph{PNAS}, 97\penalty0 (21):\penalty0 11149--11152, 2000.

\bibitem[Barab\'asi and Albert(1999)]{BarAlb99}
A.~Barab\'asi and R.~Albert.
\newblock Emergence of scaling in random networks.
\newblock \emph{Science}, 286:\penalty0 509--512, 1999.

\bibitem[Caldarelli et~al.(2002)Caldarelli, Capocci, Rios, and
  noz]{CalCapLos02}
G.~Caldarelli, A.~Capocci, P.~De~Los Rios, and M.~A.~Mu\ noz.
\newblock Scale-free networks from varying vertex intrinsic fitness.
\newblock \emph{Phys. Rev. Lett.}, 89\penalty0 (25):\penalty0 258702, 2002.
\newblock \doi{10.1103/PhysRevLett.89.258702}.

\bibitem[{de Solla Price}(1976)]{Pri76}
D.J. {de Solla Price}.
\newblock A general theory of bibliometric and other cumulative advantage
  processes.
\newblock \emph{Journal of the American Society for Information Science},
  27:\penalty0 292--306, 1976.

\bibitem[Flake et~al.(2002)Flake, Lawrence, Giles, and Coetzee]{FlaLawGil02}
G.~W. Flake, S.~R. Lawrence, C.~L. Giles, and F.~M. Coetzee.
\newblock Self-organization and identification of web communities.
\newblock \emph{IEEE Computer}, 35\penalty0 (3):\penalty0 66--71, 2002.

\bibitem[Jeong et~al.(2001)Jeong, Mason, {Barab{\'a}si}, and
  Oltvai]{JeoMasBar01}
H.~Jeong, S.P. Mason, A.-L. {Barab{\'a}si}, and Z.N. Oltvai.
\newblock Lethality and centrality in protein networks.
\newblock \emph{Nature}, 411:\penalty0 41--42, 2001.

\bibitem[Kleinberg et~al.(1999)Kleinberg, Kumar, Raghavan, Rajagopalan, and
  Tomkins]{KleKumRag99}
J.~Kleinberg, S.~R. Kumar, P.~Raghavan, S.~Rajagopalan, and A.~Tomkins.
\newblock The web as a graph: Measurements, models and methods.
\newblock In T.~Asano, H.~Imai, D.~T. Lee, S.-I. Nakano, and T.~Tokuyama,
  editors, \emph{Proceedings of the 5th Annual International Conference on
  Computing and Combinatorics, COCOON'99}, volume 1627 of \emph{LNCS}, pages
  1--17, Berlin, 1999. Springer.

\bibitem[Krings et~al.(2009)Krings, Calabrese, Ratti, and Blondel]{KriCalRat09}
G.~Krings, F.~Calabrese, C.~Ratti, and V.~D. Blondel.
\newblock Urban gravity: a model for inter-city telecommunication flows.
\newblock \emph{J.~Stat.~Mech.}, \penalty0 (7):\penalty0 L07003+8, 2009.
\newblock \doi{10.1088/1742-5468/2009/07/L07003}.

\bibitem[Lambiotte et~al.(2008)Lambiotte, Blondel, {de Kerchove}, Huens,
  Prieur, Smoreda, and {Van Dooren}]{LamBloKer08}
R.~Lambiotte, V.~D. Blondel, C.~{de Kerchove}, E.~Huens, C.~Prieur, Z.~Smoreda,
  and P.~{Van Dooren}.
\newblock Geographical dispersal of mobile communication networks.
\newblock \emph{Physica A: Statistical Mechanics and its Applications},
  387\penalty0 (21):\penalty0 5317--5325, 2008.
\newblock \doi{10.1016/j.physa.2008.05.014}.

\bibitem[Marlow et~al.(2006)Marlow, Naaman, Boyd, and Davis]{MarNaaBoy06}
C.~Marlow, M.~Naaman, D.~Boyd, and M.~Davis.
\newblock {HT06}, tagging paper, taxonomy, {Flickr}, academic article, to read.
\newblock In \emph{Proc.~of the 17th conference on Hypertext and hypermedia},
  pages 31--40. ACM, 2006.

\bibitem[McPherson et~al.(2001)McPherson, Smith-Lovin, and Cook]{PheSmiCoo01}
M.~McPherson, L.~Smith-Lovin, and J.~M. Cook.
\newblock Birds of a feather: Homophily in social networks.
\newblock \emph{Annual Review of Sociology}, 27:\penalty0 415--444, 2001.
\newblock \doi{10.1146/annurev.soc.27.1.415}.

\bibitem[Newman(2001)]{New01}
M.~E.~J. Newman.
\newblock The structure of scientific collaboration networks.
\newblock \emph{Proc. Natl. Acad. Sci.}, 98\penalty0 (2):\penalty0 404--409,
  2001.
\newblock \doi{10.1073/pnas.021544898}.

\bibitem[Pastor-Satorras and Vespignani(2001)]{PasVes01}
R.~Pastor-Satorras and A.~Vespignani.
\newblock Epidemic spreading in scale-free networks.
\newblock \emph{Physical Review Letters}, 86:\penalty0 3200--3203, 2001.
\newblock \doi{10.1103/PhysRevLett.86.3200}.

\bibitem[Schifanella et~al.(2010)Schifanella, Barrat, Cattuto, Markines, and
  Menczer]{SchBarCat10}
R.~Schifanella, A.~Barrat, C.~Cattuto, B.~Markines, and F.~Menczer.
\newblock Folks in folksonomies: social link prediction from shared metadata.
\newblock In \emph{WSDM}, pages 271--280, 2010.
\newblock \doi{10.1145/1718487.1718521}.

\bibitem[Servedio et~al.(2004)Servedio, Caldarelli, and Butt\`a]{SerCalBut04}
V.~D.~P. Servedio, G.~Caldarelli, and P.~Butt\`a.
\newblock Vertex intrinsic fitness: How to produce arbitrary scale-free
  networks.
\newblock \emph{Phys. Rev. E}, 70\penalty0 (5):\penalty0 056126, 2004.
\newblock \doi{10.1103/PhysRevE.70.056126}.

\end{thebibliography}

\end{document}